\documentclass[twocolumn,twoside,slac_two]{revtex4}
\usepackage{graphicx}
\usepackage{fancyhdr}
\def\ne{N$_{e}$}
\def\nmu{N$_\mu$}

\def\sigmappt{$\sigma_{pp}^{\it tot}$}
\def\sigmapairin{$\sigma_{p-{\rm air}}^{\it inel}$}

\def\Lambdaobsexp{$\lambda_{obs}^{exp}$}
\def\Lambdaintexp{$\lambda_{int}^{exp}$}

\begin{document}

\title{The proton-air inelastic cross-section measurement at $\sqrt{s} \approx $ 2 TeV from EAS-TOP experiment}

%
\author{G.~C.~Trinchero}
\affiliation {Istituto di Fisica dello Spazio Interplanetario (INAF), I-10133 Torino, Italy}
\affiliation {Istituto Nazionale di Fisica Nucleare, I-10125 Torino, Italy}
\author{M.~Aglietta}
\affiliation {Istituto di Fisica dello Spazio Interplanetario (INAF), I-10133 Torino, Italy}
\affiliation {Istituto Nazionale di Fisica Nucleare, I-10125 Torino, Italy}
\author{B.~Alessandro}
\affiliation {Istituto Nazionale di Fisica Nucleare, I-10125 Torino, Italy}
\author{P.~Antonioli}
\affiliation {Istituto Nazionale di Fisica Nucleare, I-40126 Bologna, Italy}
\author{F.~Arneodo}
\affiliation {Laboratori Nazionali del Gran Sasso, INFN, I-67010 Assergi (AQ), Italy}
\author{L.~Bergamasco}
\author{M.~Bertaina}
\affiliation {Dipartimento di Fisica Generale dell'Universit\`a and INFN, I-10125 Torino, Italy}
\author{A.~Castellina}
\affiliation {Istituto di Fisica dello Spazio Interplanetario (INAF), I-10133 Torino, Italy}
\affiliation {Istituto Nazionale di Fisica Nucleare, I-10125 Torino, Italy}
\author{E.~Cantoni}
\affiliation {Istituto di Fisica dello Spazio Interplanetario (INAF), I-10133 Torino, Italy}
\affiliation {Istituto Nazionale di Fisica Nucleare, I-10125 Torino, Italy}
\author{A.~Chiavassa}
\affiliation {Dipartimento di Fisica Generale dell'Universit\`a and INFN, I-10125 Torino, Italy}
\author{B.~D'Ettorre~Piazzoli}
\author{G.~Di~Sciascio}
\altaffiliation{Present address: Istituto Nazionale di Fisica Nucleare, Tor Vergata, I-00133 Roma, Italy}
\affiliation{Dipartimento di Scienze Fisiche dell'Universit\`a and INFN, I-80126 Napoli, Italy}
\author{W.~Fulgione}
\affiliation {Istituto di Fisica dello Spazio Interplanetario (INAF), I-10133 Torino, Italy}
\affiliation {Istituto Nazionale di Fisica Nucleare, I-10125 Torino, Italy}
\author{P.~Galeotti}
\affiliation {Dipartimento di Fisica Generale dell'Universit\`a and INFN, I-10125 Torino, Italy}
\author{P.~L.~Ghia}
\altaffiliation{Present address: Laboratoire de Physique Nucleaire et de Hautes Energies, Universit\'es Paris 6 et Paris 7, CNRS-IN2P3, France}
\affiliation {Istituto di Fisica dello Spazio Interplanetario (INAF), I-10133 Torino, Italy}
\affiliation {Istituto Nazionale di Fisica Nucleare, I-10125 Torino, Italy}
\author{M.~Iacovacci}
\affiliation{Dipartimento di Scienze Fisiche dell'Universit\`a and INFN, I-80126 Napoli, Italy}
\author{G.~Mannocchi}
\affiliation {Istituto di Fisica dello Spazio Interplanetario (INAF), I-10133 Torino, Italy}
\affiliation {Istituto Nazionale di Fisica Nucleare, I-10125 Torino, Italy}
\author{C.~Morello}
\affiliation {Istituto di Fisica dello Spazio Interplanetario (INAF), I-10133 Torino, Italy}
\affiliation {Istituto Nazionale di Fisica Nucleare, I-10125 Torino, Italy}
\author{G.~Navarra}
\altaffiliation{Deceased}
\affiliation {Dipartimento di Fisica Generale dell'Universit\`a and INFN, I-10125 Torino, Italy}
\author{O.~Saavedra}
\affiliation {Dipartimento di Fisica Generale dell'Universit\`a and INFN, I-10125 Torino, Italy}
\author{P.~Vallania}
\affiliation {Istituto di Fisica dello Spazio Interplanetario (INAF), I-10133 Torino, Italy}
\affiliation {Istituto Nazionale di Fisica Nucleare, I-10125 Torino, Italy}
\author{S.~Vernetto}
\affiliation {Istituto di Fisica dello Spazio Interplanetario (INAF), I-10133 Torino, Italy}
\affiliation {Istituto Nazionale di Fisica Nucleare, I-10125 Torino, Italy}
\author{C.~Vigorito}
\affiliation {Dipartimento di Fisica Generale dell'Universit\`a and INFN, I-10125 Torino, Italy}
\collaboration{EAS-TOP Collaboration}
\noaffiliation
%

\begin{abstract}
The proton-air inelastic cross section value \sigmapairin=338$\pm$21({\it stat})$\pm$19({\it syst})-28({\it syst}) mb at $\sqrt{s} \approx $ 2 TeV has been measured  by the EAS-TOP Extensive Air Shower experiment.  The absorption length of cosmic ray proton primaries cascades reaching the  maximum development at the observation level is obtained from the flux attenuation for different zenith angles (i.e. atmospheric depths). The  analysis, including the effects of the heavier primaries contribution and
systematic uncertainties, is described. The experimental result is compared with different high energy interaction models and  the relationships with the  {\it pp} ($\bar pp$) total cross section measurements are discussed.

\end{abstract}

\maketitle

\thispagestyle{fancy}

\section{Introduction}                
This energy region  $\sqrt{s} \approx $ 2 TeV is of particular relevance  because of  high energy physics and astrophysics issues.

The {\it pp} total cross section, \sigmappt, and \sigmapairin ~are related and can be inferred from each other  by means of the Glauber  theory. The whole procedure is model dependent, the  results  \cite{gaisser,d&p,kopel,block06,bell} differing of about
20\%  for $\sqrt{s}$ values in the TeV energy range. Available accelerators' measurements  at the highest energies, are themselves affected by systematic uncertainties of difficult evaluation. The  {\it pp}  ($\bar pp$) cross section measurements at  energies of $\sqrt{s} = $ 1.8 TeV \cite{cdf,d0,d01} differ of about 10\%, exceeding the statistical uncertainties of the measurements. It is therefore of primary interest to have experimental measurements of  \sigmapairin ~and \sigmappt ~at the same center of mass energy, i.e. around $\sqrt{s} \approx $ 2 TeV, at which collider data are  available.

 The interpretation of Extensive Air Shower measurements relies on simulations that use hadronic interaction models based  on theoretically guided extrapolations of the accelerator data obtained at lower energies and restricted to limited kinematical regions.
A  \sigmapairin \space direct measurement  and the  comparison of observables  as obtained from measurements and model based simulations, in the same conditions, is therefore highly recommended in order to confirm the validity of the whole analysis procedure.

Measurements of the {\it p}-air inelastic cross section performed in EAS have been reported.
Since air shower detectors cannot observe the depth of the first interaction of the primary particle, indirect methods have to be used.
Two main techniques are used: the constant  \ne-\nmu ~cuts \cite{akeno2,eastop1,argo} by means of particle arrays, and the study of the shower longitudinal profiles using fluorescence detectors \cite{fly,hires} at higher energies. 

Following the particle array technique, the primary energy is first selected by means of the muon number (\nmu). Proton induced showers at the same development stage are then selected by means of the shower size (\ne). 
The cross section of  primary particles is obtained by studying the absorption in the atmosphere ($ \lambda_{obs}$) of such showers, through their angular distribution at the observation level. 
The rate of showers
decreases exponentially  with zenith angle $\theta$ (i.e. atmospheric depth of first interaction) as:
\begin{equation}
f(\theta)= G(\theta) f(0)
\exp [- x_0 (\sec \theta - 1 )/ \lambda _{\rm obs}]
\label{flu}
\end{equation}
where  $x_{\rm 0}$ is the vertical atmospheric depth of the detector, and $ G(\theta)$ the 
angular acceptance.

With air fluorescence detectors, the absorption length $ \lambda_{obs}$ is obtained fitting the  atmospheric depth of the maximum shower development stage ($X_{max}$) distribution tail.

The observed absorption length in both cases is affected by the fluctuations in the longitudinal development of the cascades and in the detector response. Such fluctuations can be studied through simulations, providing the conversion factor {\it k} between the observed absorption length and the interaction length of  primary protons ({\it k}=$\lambda_{obs}/  \lambda_{int} $).
This factor is then 
used to convert the observed experimental absorption length \Lambdaobsexp ~into the 
interaction one \Lambdaintexp.

In this  paper we will report on the measurement of the {\it p}-air inelastic cross section  at primary energy $E_0 \approx 2 \cdot 10^{15}$ eV (i.e. $E_0~=~(1.5~\div~2.5)\cdot 10^{15}$~eV$, \sqrt{s} \approx $ 2 TeV) with the EAS-TOP  experiment. Primary energies are  below the steepening (\emph{knee}) of the primary spectrum (i.e. E$_0 < 3\cdot 10^{15}$ eV) above  which the proton flux is strongly reduced.
The  constant \ne-\nmu ~analysis has been optimized  \cite{eastop1} selecting showers at the maximum development stage where fluctuations are lower and heavier primaries rejection is improved by the request of the highest \ne values at a given primary energy.  

\begin{figure}[h]
\vspace{-5mm}
	\includegraphics[width=80mm] {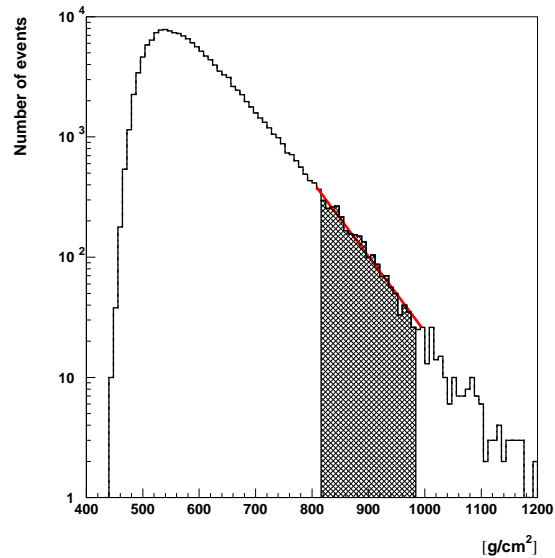}
   \caption{\label{fig:Xmax} Depth of shower maximum ($X_{max}$) distribution  for proton showers simulated with  QGSJET~II  in the selected energy range ($E_0~=~(1.5~\div~2.5)\cdot 10^{15}$ eV)
. The shaded area represents the  interval of atmospheric depths (i.e.~$1.0\leq \sec \theta \leq
1.2$) 
 considered in the analysis.}
\vspace{-4mm}
\end{figure}

The constant \ne-\nmu ~method with the selection of cascades' maximum developement stage is  equivalent to the study  of $X_{max}$ distribution tail.  As shown in Fig.~\ref{fig:Xmax}, the accessible part of the  $X_{max}$ distribution tail depends from the  vertical atmospheric depth ($x_{\rm 0}$) of the detector. This is a  limitation on  the possibility to exploit this method at different atmospheric depths and on the maximum zenith angle $\theta$ (i.e. atmospheric depth) that can be considered in the analysis without running out of  statistics.
Systematic uncertainties of the measurement and the effect of  possible contributions of heavier primaries are discussed and evaluated.  



\section{The experiment and the simulation}

The EAS-TOP array was located at Campo Imperatore, Gran
Sasso National Laboratory, 2005 m a.s.l., 820~g$/$cm$^{2}$
atmospheric depth.

 The e.m. detector was made of 35 modules $10 \
\mathrm{m^2}$ each of plastic scintillators,  4 cm thick,  distributed over an
area of $10^5 \ \mathrm{m^2}$.
The trigger is fully efficient for N$_e >
10^5$, i.e. for primary proton energies  $E_0 > 3 \cdot 10^{14}$ eV. 
The experimental resolutions  for N$_e > 2 \cdot 10^{5}$ 
are: $\sigma_{N_e}/$N$_e
\simeq 0.1$; $\sigma_{X_c} = \sigma_{Y_c} \simeq 5 \ \mathrm{m}$;
$\sigma_s \simeq 0.1$. The arrival direction of the shower is
measured from the times of flight among the modules with
resolution $\sigma_{\theta} \simeq 0.9^o$. 

The muon-hadron detector (MHD), located at one edge of the e.m. array,
 is used, for the present analysis, as a
tracking module with 9 active planes. Each plane includes two layers
of streamer tubes ($12 \ \mathrm{m}$ length, $3\times 3 \
\mathrm{cm^2}$ section) 
and is shielded by $13 \ \mathrm{cm}$ of iron. The total area of the
detector is ~$12\times12~\mathrm{m^2}$. 
A muon track is defined by the alignment of at least 6 fired wires
in different streamer tube layers leading to an  energy threshold of
$E_{\mu}^{th} \approx 1$~GeV. 

A detailed description
of the performance of the e.m. detector and of the muon-hadron detector can be found in Ref.
\cite{aglio,adinolfi}.

EAS simulations are performed with the CORSIKA program \cite{kna} with QGSJET II.03 \cite{qgsjet2}  and SIBYLL 2.1 \cite{sibyll} high energy hadronic interaction models 
These models have been widely employed for simulating atmospheric 
shower developments  and have shown to provide
consistent descriptions of different shower parameters in the considered energy range.
Hadrons with energies below  80 GeV are treated with the GHEISHA 2002 interaction model.

Proton and Helium showers have been simulated with an energy threshold 
of 10$^{15}$ eV, spectral index
$ \gamma$ = 2.7  ($\gamma _{He}  =2.65$) up to $2 \cdot 10^{16}$ eV.  KASCADE-like  spectra \cite{kascadecomp} (resulting from our own fits)  have been afterward sampled \cite{eastop1}.

Simulated events  have been analyzed using the same procedure followed for experimental data.

\section{The analysis}

The rate of showers of  given primary energy
(E$_{0,1} <$ E$_{0} <$ E$_{0,2} $) selected through their
muon number N$_\mu$ (N$_{\mu,1} <$ N$_{\mu} <$ N$_{\mu,2} $) 
and shower size N$_e$  
corresponding to maximum development (N$_{e,1} <$ N$_{e} <$ N$_{e,2} $)
is measured.

The physical quantities required for the analysis are obtained through simulations, based on QGSJET II and SIBYLL interaction models, as described below. 

Events in the specific proton primary energy range ($E_0 = (1.5 \div 2.5)\cdot 10^{15}$~eV) are selected
by means of a matrix of minimum (N$_{\mu,1}$) and maximum (N$_{\mu,2}$)
detected muon numbers for every possible combination of zenith angle 
and core distance from the  muon detector.
The selection table is obtained from simulated data
for 5 m bins in core distance (50 m $  \leq$ r
$\leq$ 100 m) and 0.025 $\sec\theta$ bins ($1.0\leq \sec \theta \leq
1.2$) for zenith angle.
 The selection of proton initiated cascades near maximum development is based on simulated distributions of the shower sizes at maximum development N$_e^{max}$ in the selected energy interval.
Choosing the shower size interval $ \overline{LogN_e^{max}}\pm \sigma_{LogN_e^{max}}$ (i.e. 6.01$<$ Log N$_e$ $<$ 6.17 for both interaction models)
provides the selection of about $ 65 \%$ of the events around the maximum of the  N$_e^{max}$ distribution.\\
\begin{figure}[h]
\vspace{-5mm}
\includegraphics[width=80mm]{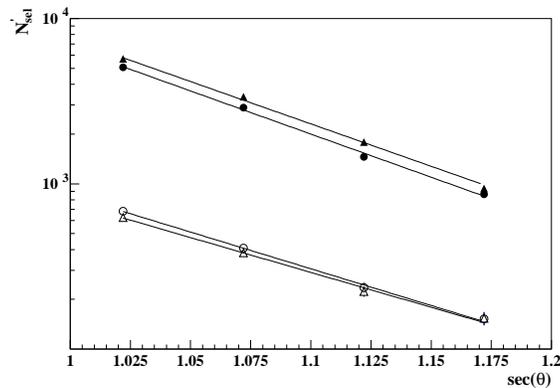}
\vspace{-6mm}    \caption{\label{fig:k} Acceptance corrected  number of events vs. sec$\theta$ for the simulated (solid circle for QGSJET II and solid triangle for SIBYLL) and experimental  data selected with the N$_{\mu}$-N$_{e}$ cuts calculated with the two interaction models (open circle for QGSJET II and open triangle for SIBYLL). 
    The fits with expression (\ref{flu}) providing the $\lambda_{obs}$ values are also shown (continuous lines).}
\end{figure}
\begin{figure}[h]
\vspace{-6mm}
\includegraphics[width=80mm]{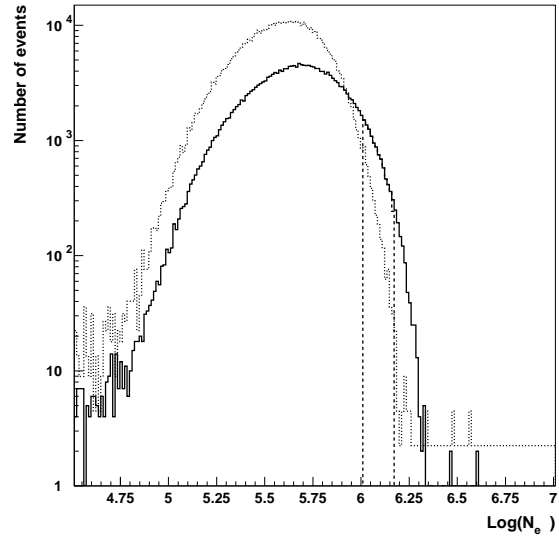}
\vspace{-4mm}
    \caption{\label{fig:Nelog} Shower size distribution (Log(N$_e$)) for proton (continuous) and helium (dotted) showers simulated with  QGSJET II  in the selected energy range 
. The two vertical dashed lines delimitate the Log(N$_e$) interval considered in the analysis.}
\vspace{-4mm}
\end{figure}

The interaction length  $  \lambda_{int}^{sim} $ is  obtained as the average proton interaction depth in the selected energy range, 
it results to be $  \lambda_{int}^{sim} = 60.3 \pm 0.1 $ g$/$cm$^2$ for QGSJET II and $\lambda_{int}^{sim} = 59.4 \pm 0.1 $ g$/$cm$^2$ for SIBYLL.

The acceptance corrected numbers of selected events $N_{sel}'$ vs. zenith angle  are shown in Fig.~\ref{fig:k}.
The fit with expression (\ref{flu}) provides  $  \lambda_{obs}^{sim} = 68.5\pm 1.4$~g$/$cm$^2$ for QGSJET II and $  \lambda_{obs}^{sim} = 69.9\pm 1.4$~g$/$cm$^2$ for SIBYLL. Therefore  {\it k} = $ \lambda_{obs}^{sim} /  \lambda_{int}^{sim} = 1.14\pm 0.02$ for QGSJET II and {\it k}~$=~1.18\pm 0.02$  for SIBYLL 

\section{Results}
The same analysis procedure discussed for the simulations is applied to the experimental data.
The corresponding event numbers as a function of sec($\theta$)
are also shown in Fig.~\ref{fig:k}, together with their fits  providing 
$\lambda_{obs}^{exp}=80.2\pm4.3~{\rm g/cm^{2}}$ and $\lambda_{obs}^{exp}=84.7\pm5.0~{\rm g/cm^{2}}$ for QGSJET~II and  SIBYLL respectively.
From the relation $\lambda_{int}^{exp}$=$\lambda_{obs}^{exp} / k $, we obtain
 $ \lambda_{int}^{exp}=\lambda_{p-\rm air} = 70.7\pm4.2~{\rm g/cm^{2}}$
 for QGSJET~II
and $\lambda_{int}^{exp}=\lambda_{p-\rm air} = 71.8\pm4.5~{\rm g/cm^{2}}$ for SIBYLL. 
The {\it p}-air inelastic cross section is then  obtained from the relation  $\sigma^{\rm inel}_{p- \rm air} ({\rm mb}) 
= 2.41\cdot 10^4/\lambda_{p-\rm air}$, and results to be $\sigma^{\rm inel}_{p- \rm air} = 341 \pm 20~  {\rm mb}$  with QGSJET~II and $\sigma^{\rm inel}_{p- \rm air} = 336 \pm 21~{\rm mb}$ with SIBYLL analysis.

The contribution of heavier nuclei  has been  evaluated by simulating helium primaries with QGSJET II, assuming the KASCADE spectrum and composition, which accounts for an helium flux about twice that of the  protons in the energy range of interest but, as shown in  Fig.~\ref{fig:Nelog},
due to the high \ne values requested in our analysis, the helium conatmination is reduced to less than 25\%. The  overall simulated observed absorption length becomes $\lambda_{obs}^{sim(p+He)}$ = $ 62.6 \pm 1.0 $ g/cm$^2$, which implies k$^{(p+He)}$ = $ 1.04 \pm 0.02 $ , and \sigmapairin 
 = 312 $ \pm 17 $~mb. Heavier primaries (i.e. CNO) hardly pass the N$_{\mu}$-N$_{e}$  cuts.

 The analysis procedure  based on one interaction model has been applied to a simulated experimental data set  produced with a different interaction model with known  p-air inelastic cross section in order to evaluate the systematic uncertainties. The result of the  data simulated with SIBYLL (\sigmapairin=406$\pm$1~mb) when analyzed with QGSJET II is \sigmapairin=393$\pm$11~mb  ($\Delta$\sigmapairin=-13$\pm$11~mb) and viceversa \sigmapairin=419$\pm$12~mb when the data simulated with  QGSJET II (\sigmapairin=400$\pm$1~mb  ($\Delta$\sigmapairin=+19$\pm$12~mb) are analyzed with SIBYLL. 

An  experimental data set has been simulated using QGSJET II with HDPM  \cite{hdpm}  cross section value (\sigmapairin=367$\pm$1~mb)
in order to check the capability to discriminate between two different values of the p-air cross section within the same interaction model.
The lower  cross section value is  clearly discriminated by the anlysis based on  SIBYLL that gives \sigmapairin=372$\pm$13~mb  ($\Delta$\sigmapairin=-5$\pm$13~mb).
The differences  between the simulated and measured values ($\Delta$\sigmapairin) are both positive and negative and compatible with the  statistical uncertainties. 
Therefore we define as maximum systematic uncertainty the value $\sigma_{syst} = 19$ $  {\rm mb}$ which provides the $\chi^2$ value corresponding to $99\%~{\rm C.L}$ for the  distribution of the four deviations $\Delta$\sigmapairin .

\begin{figure}[h]
\vspace{-5mm}
	\includegraphics[width=80mm]{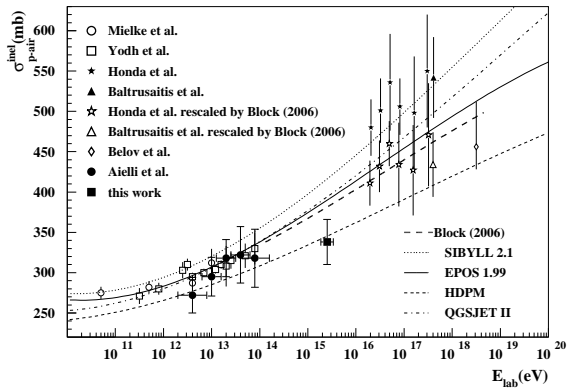}
\vspace{-4mm}
    \caption{\label{fig:inpair}{\it p}-air inelastic cross section data 
including the present measurement (solid square), and different  hadronic interaction models. 
}
\vspace{-4mm}
\end{figure}

\section{Conclusions}
Combining the results obtained with the two considered interaction models and including 
the systematic uncertainties 
the {\it p}-air inelastic cross section is: 
\vspace{2mm}

$\sigma^{\rm inel}_{p- \rm air} = 338 \pm 21_{stat} \pm 19_{syst} -29_{syst(He)}  {\rm mb}$
\vspace{2mm}

As shown in Fig.~\ref{fig:inpair}, this  value is   about 15\% smaller than the values in use within QGSJET II and SIBYLL  and in better agreement with Refs. \cite{block06,hor,hdpm}.
Predicted  $\sigma^{\rm inel}_{p- \rm air}$ values, obtained from different  \sigmappt~Tevatron measurements at $\sqrt{s} = $ 1.8 TeV by using different calculations based on the Glauber theory, are  reported in Fig.~\ref{fig:ppair}.
The present measurement is consistent with smaller values of the  $\bar pp$ total cross section (\sigmappt=72.8$\pm$3.1 mb \cite{d0}, and \sigmappt =71$\pm$2 mb \cite{d01}), and the {\it pp}  to {\it p}-air  calculations predicting for a given value of \sigmappt, a smaller value of $\sigma_{\rm in}^{p- \rm air}$ 
\cite{block06,bell}.
\begin{figure}[h]
\vspace{-1mm}
    \includegraphics[width=80mm]{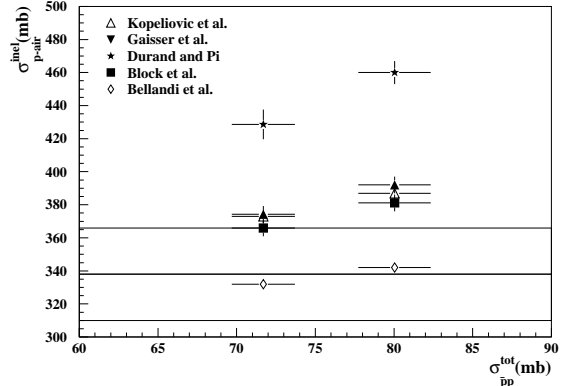}
\vspace{-4mm}
    \caption{\label{fig:ppair}  {\it p}-air inelastic  vs.  {\it $ \bar pp $} total cross section data.  The present result ($\pm$ 1 s.d., solid lines) is shown together with the results of different calculations
 derived from  {\it $ \bar pp $} measurements reported at $ \sqrt{s} = $ 1.8 TeV.  
    }
\vspace{-4mm}
    \label{fig:ppair}
\end{figure}
Independently from the cross section analysis, the measured values of the absorption length $\lambda_{obs}$  are about 15\% higher than the simulated ones for both the considered interaction models. This can be probably ascribed to the fact  that the measured cascades penetrate deeper into the atmosphere than predicted by the interaction models.

\end{document}